\documentclass[preprint,12pt]{elsarticle}

\usepackage{graphics}

\usepackage{amssymb}

\journal{Nuclear Physics A}

\begin{document}

\begin{frontmatter}



\title{Update of High Resolution $(e,e'K^+)$ Hypernuclear Spectroscopy at Jefferson Lab's Hall A}

\address[Roma]{INFN Sezione di Roma, Rome, Italy}
\address[FIU]{Florida International University, Miami, FL 32306, USA}
\address[Praha]{Nuclear Physics Institute, \v{R}e\v{z} near Prague, Czech Republic}
\address[UMD]{University of Maryland, College Park, MD, Czech Republic}
\address[ISS]{INFN gr. Sanit\`a coll. Sezione di Roma and Istituto Superiore di Sanit\`a, Rome, Italy}
\address[JLab]{Thomas Jefferson National Accelerator Facility, Newport News, VA 23606, USA}
\address[Bari]{INFN Sezione di Bari and Dipartimento di Fisica, Bari, Italy}
\address[Tre]{INFN Sezione di Roma Tre, Rome, Italy}

\author[Roma]{F.~Cusanno \corref{cor1}}
\ead{cusanno@jlab.org, francesco.cusanno@ph.tum.de}
\author[FIU]{A.~Acha}
\author[Praha]{P.~Byd\v{z}ovsk\'y}
\author[UMD]{C.~C.~Chang}
\author[ISS]{E.~Cisbani}
\author[JLab]{C.~W.~De Jager}
\author[Bari]{R.~De~Leo}
\author[ISS]{S.~Frullani}
\author[Roma,ISS]{F.~Garibaldi}
\author[JLab]{D.~W.~Higinbotham}
\author[Tre]{M.~Iodice}
\author[JLab]{J.~J.~LeRose}
\author[FIU]{P.~Markowitz}
\author[Bari]{S.~Marrone}
\author[Praha]{M.~Sotona}
\author[Roma]{G.~M.~Urciuoli}
\author{\\ for the Hall A Collaboration}

\cortext[cor1]{Corresponding author. Presently at E12 Physik Department and Excellence Cluster 'Universe', Technische Universit\"at M\"unchen, Garching b. M\"unchen, Germany.}

\begin{abstract}
Updated results of the experiment E94-107 hypernuclear spectroscopy in Hall A of the Thomas Jefferson National Accelerator Facility (Jefferson Lab), are presented. The experiment provides high resolution spectra of excitation energy for $^{12}_{\Lambda}B$, $^{16}_{\Lambda}N$, and $^{9}_{\Lambda}Li$ hypernuclei obtained by electroproduction of strangeness. A new theoretical calculation for $^{12}_{\Lambda}B$, final results for $^{16}_{\Lambda}N$, and discussion of the preliminary results of $^{9}_{\Lambda}Li$ are reported. 
\end{abstract}


\end{frontmatter}


\section{Introduction}
\label{Intro}

The investigation of hypernuclei using electromagnetic probes presents the following advantages:
\begin{itemize}
\item[-] energy resolution of the order of few hundreds keV is obtained;  
\item[-] natural and unnatural parity states can be populated in the produced hypernucleus, due to the large 3-momentum transfer (($\vec{q}$~$\gtrsim$ 250~MeV/c) and strong spin flip;
\item[-] {\it mirror} hypernuclei are produced with respect to those investigated with hadron probes;
\end{itemize}
Despite the very small cross sections of electroproduction of hypernuclei, Jefferson Lab is able to provide high-current and high quality beam for successfully performing such experiments. For this purpose, a pair of septum magnets and a RICH detector \cite{LeRose} were added to the Hall A standard apparatus \cite{NIMA}. 

\section{Results for $^{12}C(e,e'K^+)^{12}_{\Lambda}B$}
\label{Carb}

\begin{figure}[h!]
\centering
\includegraphics[width=9.5cm, angle=0]{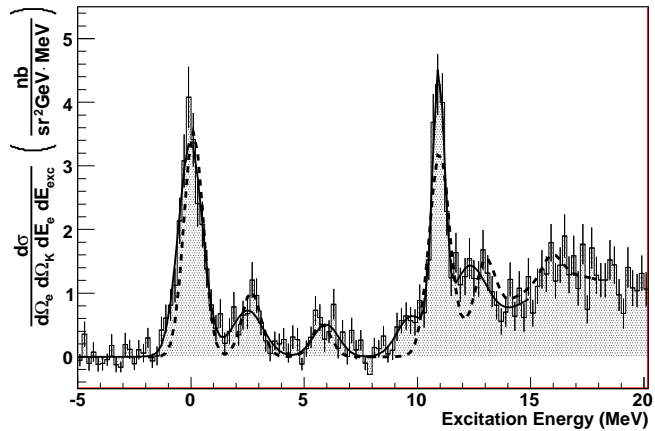}
\caption{Excitation energy spectrum of the 
$^{12}C(e,e'K^+)^{12}_{\Lambda}B$ compared with a new theoretical calculation (dashed line, see text).}
\label{fig:newcarb}
\end{figure}

The results of the investigation of the 
$^{12}C(e,e'K^+)^{12}_{\Lambda}B$ have been already described elsewhere \cite{Mauro}. With respect to the previously published data, a new theoretical prediction has been calculated and its comparison with the experimental spectrum is shown in 
Fig.~\ref{fig:newcarb}. The theoretical cross sections, for all of the investigated hypernuclei, are obtained in the framework of the distorted wave impulse approximation (DWIA) \cite{DWIA} using the Saclay-Lyon (SLA) model \cite{SLA} for the elementary $p(e,e'K^+)\Lambda$ reaction. Shell-model wave functions are determined using a parametrization of the $\Lambda N$ interaction that fits the precise $\gamma$-ray hypernuclear spectra of $^{7}_{\Lambda}Li$ \cite{Ukai}.
The new curve has been obtained with an improved calculation of the optical potential of the $K^+$, consisting of a stronger absorbtion with respect to the previous predictions. 
Now the agreement with the data for the ground state and the core-excited states is very good, while the discrepancy for the p-shell part of the spectrum is slightly increased with respect to the previous calculations.

\section{Results for $^{16}O(e,e'K^+)^{16}_{\Lambda}N$}
\label{Oxy}
\begin{figure}
\centering
\includegraphics[width=9.5cm, angle=0]{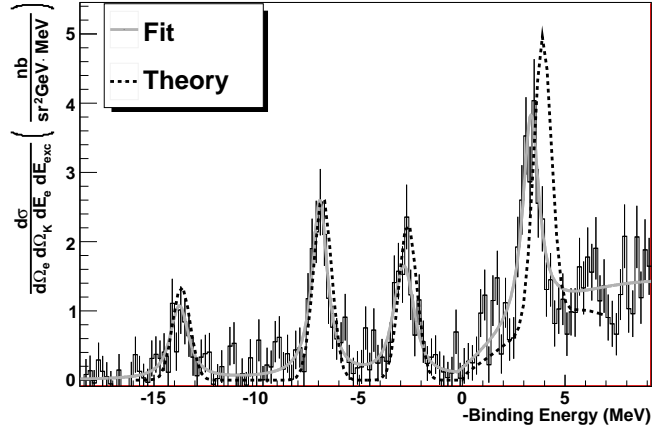}
\caption{Excitation energy spectrum of the $^{16}O(e,e'K^+)^{16}_{\Lambda}N$ compared with the theoretical calculation.}
\label{fig:oxy}
\end{figure}
The results of the investigation of the 
$^{16}O(e,e'K^+)^{16}_{\Lambda}N$ have been recently published elsewhere \cite{OxyPaper}. They are briefly summarized below:
\begin{itemize}
\item[-] for the first time, the energy spectrum of $^{16}_{\Lambda}N$ has been obtained, allowing a direct comparison with the available data on the mirror $^{16}_{\Lambda}O$ hypernucleus;  
\item[-] the use of a waterfall target \cite{waterfall} allowed the precise determination of the binding energy, calibrated using the reaction of elementary $p(e,e'K^+)\Lambda$ reaction on hydrogen;
\item[-] the comparison with the theoretical predictions provide restrictions on the spacings of the levels contributing to each peak and, therefore, on the $\Lambda$ spin-orbits part of the effective interaction $s_\Lambda$;
\end{itemize}
The Table \ref{tab:results} and Fig.~\ref{fig:oxy} report the details of the comparison of the fit to the data with the theoretical calculations.
\begin{table*}[ht!]
\caption{Levels and cross sections obtained by fitting the
$^{16}$O$(e,e'K^+)^{16}_{\Lambda}N$ spectrum (first two columns) compared with
theoretical predictions (last four columns).
\label{tab:results}}
\begin{tabular}{ccccccc}
\hline
$E_x$ &  Cross section & $E_x$  & Wave
function & $J^\pi$ & Cross section \\
 (MeV) & $(nb/sr^2/GeV)$ & (MeV) &   &   &
 $(nb/sr^2/GeV)$\\
 \hline
 \phantom{1}0.00\phantom{ $\pm$ 0.06}  & 1.45 $\pm$ 0.26 & 0.00 
& $p^{-1}_{1/2}\otimes s_{1/2\Lambda}$& $0^-$ & 0.002 \\
   &  & 0.03 & $p^{-1}_{1/2}\otimes s_{1/2\Lambda}$ & $1^-$ & 1.45 \\
 & & & & & & \\
\phantom{1}6.83 $\pm$ 0.06  & 3.16 $\pm$ 0.35 & 6.71 &
$p^{-1}_{3/2}\otimes s_{1/2\Lambda}$ & $1^-$ & 0.80 \\
   & & 6.93 &$p^{-1}_{3/2}\otimes s_{1/2\Lambda}$ & $2^-$   &2.11 \\
 & & & & & & \\
10.92 $\pm$ 0.07  & 2.11 $\pm$ 0.37 & 11.00 &
$p^{-1}_{1/2}\otimes p_{3/2\Lambda}$ & $2^+$ & 1.82 \\
  & & 11.07 & $p^{-1}_{1/2}\otimes p_{1/2\Lambda}$ & $1^+$ & 0.62 \\
 & & & & & & \\
17.10 $\pm$ 0.07 & 3.44 $\pm$ 0.52 & 17.56 & 
$p^{-1}_{3/2}\otimes p_{1/2\Lambda}$ & $2^+$ & 2.10 \\
  & & 17.57 &
$p^{-1}_{3/2}\otimes p_{3/2\Lambda}$ & $3^+$ & 2.26 \\
\hline
\end{tabular}
\end{table*}
Four distinguished peaks are observed.  
The experimental results show a good agreement with the predictions. The $\Lambda$ separation energy of the ground state results in
$B_{\Lambda}=13.76 \pm 0.16$ MeV. 
The largest discrepancy between theory and experiment is in the position of the fourth peak, but the prediction of the excitation energies of the positive-parity states (see Table \ref{tab:results}) can be uncertain by few hundreds keV \cite{OxyPaper}. 

\section{Results for $^{9}Be(e,e'K^+)^{9}_{\Lambda}Li$}
\label{Ber}

\begin{figure}[htb!]
\centering
\includegraphics[width=9.5cm, angle=0]{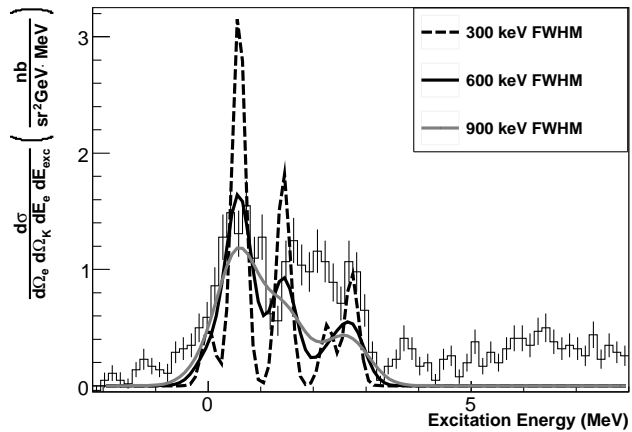}
\caption{Excitation energy spectrum of the $^{9}Be(e,e'K^+)^{9}_{\Lambda}Li$ compared with theoretical curves assuming a FWHM spread of the levels of 300~keV, 600~keV, and 900~keV.}
\label{fig:be1}
\end{figure}

The results of the analysis of the $^{9}Be(e,e'K^+)^{9}_{\Lambda}Li$, reported in Fig.~\ref{fig:be1}, are still preliminary. Fine corrections to the excitation energy spectrum are still ongoing, but different interpretations can already be argued on the present status of the analysis.\\

\begin{figure}[htb!]
\centering
\includegraphics[width=9.5cm, angle=0]{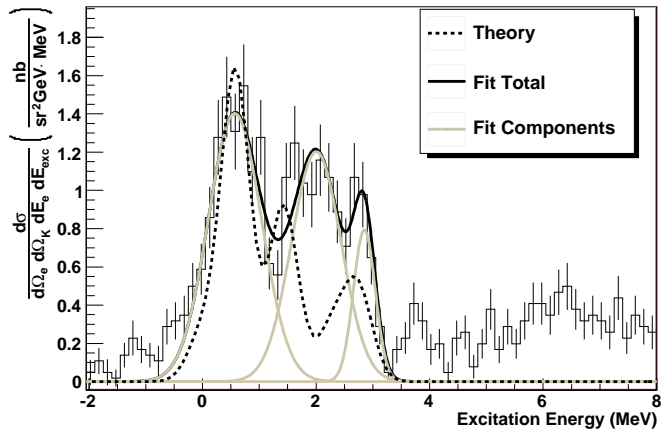}
\caption{The spectrum shown in Fig.~\ref{fig:be1} fitted with three peaks.}
\label{fig:be2}
\end{figure}

\begin{figure}[bth!]
\centering
\includegraphics[width=9.5cm, angle=0]{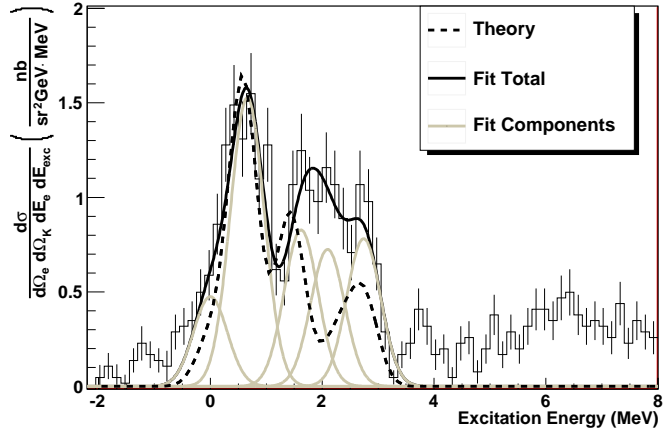}
\caption{Excitation energy spectrum of the $^{9}Be(e,e'K^+)^{9}_{\Lambda}Li$ superimposed with a theoretical curve and compared with a fit assuming five peaks of the same width in region of interest.}
\label{fig:be3}
\end{figure}

\begin{table*}[h!]
\centering
\caption{Preliminary results of levels obtained by fitting the
$^{9}$Be$(e,e'K^+)^{9}_{\Lambda}Li$ spectrum assuming three peaks of different widths.
\label{tab:threep}}
\begin{tabular}{ccc}
\hline
$E_x$ &  Cross section & Resolution FWHM  \\
 (MeV) & $(nb/sr^2/GeV)$ & (MeV)\\
 \hline
 $0.0 \pm 0.04$  & $1.25 \pm 0.15$ & $1.10 \pm 0.18$ \\
 & & \\
1.39 $\pm 0.07$  & $1.12 \pm 0.15$ & $1.02 \pm 0.21$  \\
 & &  \\
2.25 $\pm 0.05$  & $0.38 \pm 0.11$ & $0.44 \pm 0.12$ \\
\hline
\end{tabular}
\end{table*}
For this hypernucleus, it is not possible to perform a peak search with the algorithm used for the other spectra, because a convolution of peaks is clearly observed.
The bin for the histogram of the excitation energy spectrum is set to 150~keV.
In Fig.~\ref{fig:be1}, the theoretical curve is superimposed on the data. Different energy resolutions have been assumed to spread the levels. In the case of 600~keV resolution, it clearly results in three main structures also compatible with the convolutions of multiple levels in the measured spectrum.
If a fit with only two peaks is performed, the FWHM widths of the peaks are 1.0~MeV and 1.7~MeV respectively. The reduced 
$\chi^2$ ($\chi^2/NDF$) for the fit is 1.40. 
The fit procedure is described elsewhere \cite{Mauro}.
A fit with three peaks is also guided by the three main structures predicted by the theory. In this case the results are reported in 
Table \ref{tab:threep} and Fig.~\ref{fig:be2}. 
The value of $\chi^2/NDF$ for the fit is 1.34.\\
Finally, an attempt to disentangle the different contributions coming from the multiple levels calls for a fit with five peaks, as guided by the theoretical model. Since in principle the width of the peaks should be the same for the single states, determined by the experimental resolution, in this case the fit is performed assuming the same resolution for the single states. The result of this fit for the width of the individual peaks is $710\pm 140$~keV FWHM, in agreement with the measurements on $^{12}_{\Lambda}B$ and $^{16}_{\Lambda}N$.
The result is shown in Fig.~\ref{fig:be3}.
The value of $\chi^2/NDF$ for the fit is 1.14.  \\

In any case, the discrepancy between the experimental data and the theoretical curve is rather evident. 
The preliminary comparison is shown in the Table \ref{tab:be}. 
\begin{table*}[hb!]
\centering
\caption{Preliminary results of levels and cross sections obtained by fitting the
$^{9}Be(e,e'K^+)^{9}_{\Lambda}Li$ spectrum with five peaks compared with theoretical predictions.
\label{tab:be}
}
\begin{tabular}{ccccc}
\hline
$E_x$ &  Cross section & $E_x$  & $J^\pi$ & Cross section \\
 (MeV) & $(nb/sr^2/GeV)$ & (MeV) &  &
 $(nb/sr^2/GeV)$\\
 \hline
 0.00\phantom{ $\pm$ 0.24}  & 0.25 $\pm$ 0.13 & 0 & $3/2^+$ & 0.159 \\
 & & & & \\
0.73 $\pm$ 0.07  & 1.02 $\pm$ 0.25 & 0.58 & $5/2^+$ & 1.04 \\
 & & & & \\
1.73 $\pm$ 0.34  & 0.45 $\pm$ 0.15 & 1.43 & $1/2^+$, $3/2^+$ & 0.591 
\\
 & & & & \\
2.12 $\pm$ 0.11 & 0.41 $\pm$ 0.14 & 2.27 & $5/2^+$ & 0.169 \\
 & & & & \\
2.82 $\pm$ 0.04 & 0.48 $\pm$ 0.12 & 2.73 & $7/2^+$ & 0.311 \\
\hline
\end{tabular}
\end{table*}
However, since the individual peaks in the observed structure are not resolved, small corrections expected in the final results might also play a important role in better defining the spectrum.
Furthermore, precise determination of the binding energy might be possible using the data obtained from the Beryllium windows in the waterfall target, or using the better known binding energy of the $^{12}_{\Lambda}B$, measured during the same beam time with the same setup.  
\section{Conclusion}
\label{concl}
Experiment E94-107 at Jefferson Lab completed very successful runs on three different targets. The results of the hypernuclear spectroscopy performed on $^{12}C$ target and $^{16}O$ targets are published and they provide important elements for a better understanding of the strangeness physics. Preliminary results on the $^{9}Be$ target, where the spin-spin term of the $\Lambda$-N potential can also be investigated provide additional interesting data for comparison with the theoretical models.  
\\

\textbf{Acknowledgements}
\\
This work was supported by US Department of Energy contract DE-AC05-84ER40150 Modification No. M175 under which formerly the Southeastern Universities Research Association and presently the Jefferson Science Associates LLC operates the Thomas Jefferson National Accelerator Facility, by the Italian Istituto Nazionale di Fisica Nucleare (INFN), by the US Department of Energy under contracts W-31-109-ENG-38, DE-FG02-99ER41110, and DE-AC02-98-CH10886, by the Grant Agency of Czech Republic No. 202/08/0984, and by the French CEA and CNRS/IN2P3. \\
One of the authors (F. Cusanno) is grateful to the Excellence Cluster 'Universe', Germany, and to the Organizers of the Hyp-X Conference for their support.
\\

\textbf{References}






\end{document}